\documentclass[11pt,preprint]{aastex}
\usepackage{psfig}

\slugcomment{Accepted by the Astrophysical Journal}
 
\shorttitle{Four Highly Dispersed MSPs from the PALFA Survey}

\shortauthors{Crawford et al.} 
 
\begin{document}
 
\title{Four Highly Dispersed Millisecond Pulsars Discovered in the
Arecibo PALFA Galactic Plane Survey}
 
\author{F. Crawford\altaffilmark{1},
K. Stovall\altaffilmark{2,3},
A. G. Lyne\altaffilmark{4}, 
B. W. Stappers\altaffilmark{4}, 
D. J. Nice\altaffilmark{5},
I. H. Stairs\altaffilmark{6},
P. Lazarus\altaffilmark{7,8},
J. W. T. Hessels\altaffilmark{9,10}, 
P. C. C. Freire\altaffilmark{8}, 
B. Allen\altaffilmark{11,12}, 
N. D. R. Bhat\altaffilmark{13}, 
S. Bogdanov\altaffilmark{14}, 
A. Brazier\altaffilmark{15}, 
F. Camilo\altaffilmark{14},  
D. J. Champion\altaffilmark{8}, 
S. Chatterjee\altaffilmark{15}, 
I. Cognard\altaffilmark{16},  
J. M. Cordes\altaffilmark{15}, 
J. S. Deneva\altaffilmark{17}, 
G. Desvignes\altaffilmark{8,16},
F. A. Jenet\altaffilmark{2}, 
V. M. Kaspi\altaffilmark{7}, 
B. Knispel\altaffilmark{11}, 
M. Kramer\altaffilmark{8}, 
J. van Leeuwen\altaffilmark{9,10},  
D. R. Lorimer\altaffilmark{18}, 
R. Lynch\altaffilmark{7},
M. A. McLaughlin\altaffilmark{18}, 
S. M. Ransom\altaffilmark{19},
P. Scholz\altaffilmark{7},
X. Siemens\altaffilmark{12}, 
A. Venkataraman\altaffilmark{17}
}

\altaffiltext{1}{Department of Physics and Astronomy, Franklin and
Marshall College, P.O. Box 3003, Lancaster, PA 17604, USA; email:
fcrawfor@fandm.edu}

\altaffiltext{2}{Center for Gravitational Wave Astronomy, University
of Texas at Brownsville, Brownsville, TX 78520, USA}

\altaffiltext{3}{Department of Physics and Astronomy, University of
Texas at San Antonio, San Antonio, TX 78249, USA}

\altaffiltext{4}{Jodrell Bank Centre for Astrophysics, University of
Manchester, Manchester M13 9PL, UK}

\altaffiltext{5}{Department of Physics, Lafayette College, Easton, PA
18042, USA}

\altaffiltext{6}{Department of Physics and Astronomy, University of
British Columbia, 6224 Agricultural Road, Vancouver, BC V6T 1Z1,
Canada}

\altaffiltext{7}{Department of Physics, McGill University, 3600
University Street, Montreal, QC H3A 2T8, Canada}

\altaffiltext{8}{Max-Planck-Institut f\"{u}r Radioastronomie, auf dem
Huegel 69, 53121 Bonn, Germany}

\altaffiltext{9}{ASTRON, the Netherlands Institute for Radio
Astronomy, Postbus 2, 7990 AA, Dwingeloo, The Netherlands}

\altaffiltext{10}{Astronomical Institute ``Anton Pannekoek,''
University of Amsterdam, Science Park 904, 1098 XH Amsterdam, The
Netherlands}

\altaffiltext{11}{Albert-Einstein-Institut, Max-Planck-Institut
f\"{u}r Gravitationsphysik, D-30167 Hannover, Germany, and Institut
f\"{u}r Gravitationsphysik, Leibniz Universit\"{a}t Hannover, D-30167
Hannover, Germany}

\altaffiltext{12}{Physics Department, University of Wisconsin at
Milwaukee, Milwaukee, WI 53211, USA}

\altaffiltext{13}{Swinburne University, Center for Astrophysics and
Supercomputing, Hawthorn, Victoria 3122, Australia}

\altaffiltext{14}{Columbia Astrophysics Laboratory, Columbia
University, New York, NY 10027, USA}

\altaffiltext{15}{Astronomy Department, Cornell University, Ithaca, NY
14853, USA}

\altaffiltext{16}{Laboratoire de Physique et Chimie de l'Environnement et
de l'Espace, LPC2E, CNRS et Universit\'{e} d'Orl\'{e}ans, and Station de
radioastronomie de Nan\c{c}ay, Observatoire de Paris, France}

\altaffiltext{17}{Arecibo Observatory, HC3 Box 53995, Arecibo, PR
00612, USA}

\altaffiltext{18}{Department of Physics, West Virginia University,
Morgantown, WV 26506, USA}

\altaffiltext{19}{National Radio Astronomy Observatory, 520 Edgemont
Rd., Charlottesville, VA 22903, USA}

\begin{abstract}
We present the discovery and phase-coherent timing of four highly
dispersed millisecond pulsars (MSPs) from the Arecibo PALFA Galactic
plane survey: PSRs J1844+0115, J1850+0124, J1900+0308, and J1944+2236.
Three of the four pulsars are in binary systems with low-mass
companions, which are most likely white dwarfs, and which have orbital
periods on the order of days. The fourth pulsar is isolated. All four
pulsars have large dispersion measures (DM $> 100$ pc cm$^{-3}$), are
distant ($\ga 3.4$ kpc), faint at 1.4 GHz ($\la 0.2$ mJy), and are
fully recycled (with spin periods $P$ between 3.5 and 4.9 ms). The
three binaries also have very small orbital eccentricities, as
expected for tidally circularized, fully recycled systems with
low-mass companions.  These four pulsars have DM/$P$ ratios that are
among the highest values for field MSPs in the Galaxy.  These
discoveries bring the total number of confirmed MSPs from the PALFA
survey to fifteen.  The discovery of these MSPs illustrates the power
of PALFA for finding weak, distant MSPs at low-Galactic
latitudes. This is important for accurate estimates of the Galactic
MSP population and for the number of MSPs that the Square Kilometer
Array can be expected to detect.
\end{abstract}


\section{Introduction}

The PALFA survey is an ongoing, large-scale pulsar survey of the
Galactic plane that uses the Arecibo 305-m radio telescope and the
Arecibo L-Band Feed Array (ALFA) 7-beam multi-beam receiver
\citep{cfl+06}.  It is one of Arecibo's key science projects, and it
will ultimately cover the entire Arecibo-visible sky within
$5^{\circ}$ of the Galactic plane (longitudes of $32^{\circ} \la l \la
77^{\circ}$ and $168^{\circ} \la l \la 214^{\circ}$).  PALFA observes
at relatively high observing frequencies ($1220-1520$\,MHz) in order
to mitigate the deleterious effects that interstellar dispersion and
scattering have on the detection of distant pulsars at low Galactic
latitudes.  In this sense, the PALFA survey is similar to the highly
successful Parkes Multibeam Pulsar Survey (PMPS; Manchester et
al. 2001\nocite{mlc+01}), but with increased time and frequency
resolution, such as those of the current Parkes HTRU survey
\citep{kjv+10}.

Compared to past Arecibo surveys, PALFA explores a far larger spatial
volume due to its high time and frequency resolution, enabling
discovery of faint, highly dispersed millisecond pulsars
(MSPs)\footnote{An MSP typically refers to a (partially or fully)
recycled pulsar having a small surface magnetic field strength ($B \la
10^{10}$ G) and a large characteristic age ($\ga 10^{9}$ yr).}  in the
Galactic plane, which tend to be at larger distances.  Since the
majority of MSPs have binary companions ($\sim 80$\%, Lorimer
2008\nocite{l08}), they are often interesting test cases for studies
of exotic binary stellar evolution (e.g., Archibald et
al. 2009\nocite{asr+09}; Freire et al. 2011\nocite{fbw+11}).  The
unsurpassed sensitivity of the Arecibo telescope is highly
advantageous for detecting binary MSPs because the PALFA pointing
dwell times are only a small fraction (less than 10\%) of the orbital
periods of all known binary radio pulsars.  In this regime, linear
acceleration searches are highly effective at recovering
Doppler-smeared periodicities \citep{jk91}.

The discovery of PSR J1903+0327 by \citet{crl+08} is an excellent
illustration of the PALFA survey's sensitivity to highly dispersed
pulsars.  PSR J1903+0327 has a spin period of only 2.15 ms and has the
highest dispersion measure (DM) of all known completely recycled
Galactic MSPs, 297 pc cm$^{-3}$.  It occupies a region of DM-period
phase-space that has previously been unexplored for pulsars in the
Galactic field.  For example, of the 90 recycled Galactic field radio
pulsars currently listed in the ATNF catalog (apart from the PALFA
discoveries)\footnote{We also exclude the high-DM MSPs in globular
clusters, because these are found by targeted searches conducted with
very long integration times and at even higher observing frequencies
(generally $\sim 2$ GHz).}  with spin periods $P < 25$ ms, only 9 have
DM $> 100$ pc cm$^{-3}$, and only 14 have DM-derived distances listed
in the catalog that are greater than 3 kpc.

One of the main motivations for finding distant, highly dispersed MSPs
is to determine a more complete census of the Galactic MSP population,
which is currently biased by the large number of nearby ($\la 2$ kpc)
sources -- especially given the recent discoveries of more than 40
generally nearby MSPs through targeted searches of {\it Fermi}
gamma-ray sources (e.g., Ransom et al. 2011\nocite{rrc+11}).  MSPs are
the longest-living active manifestations of neutron stars, with active
lives hundreds to thousands of times longer than those of normal
pulsars, magnetars or accreting neutron stars.  Hence they give
valuable insight into the Galactic neutron star population and binary
stellar evolution in particular.  The planned Square Kilometer Array
(SKA; e.g., Carilli \& Rawlings 2004\nocite{cr04}) should be able to
detect a large fraction of the MSPs in the Galaxy, and finding high-DM
MSPs beforehand will help tell us how many to expect.  Furthermore,
MSPs are excellent probes of the interstellar medium, and the
discovery of more distant MSPs whose signals likely pass through
multiple scattering screens opens new possibilities (and challenges)
in this area. For instance, scattering measurements of large-DM MSPs
can be used to compare the observed effects of scattering on timing
behavior with predictions of those effects.

MSPs can also be used as high-precision astronomical clocks.  Stable
MSPs form the basis of the pulsar timing efforts of the International
Pulsar Timing Array consortium to detect long-period gravitational
radiation from observations of pulsar timing residuals
\citep{jhl+05,h++10}.  The recent precision mass measurements of PSR
J1903+0327 \citep{fbw+11} and PSR J1614$-$2230 \citep{dpr+10} indicate
that MSPs can have masses well above the Chandrasekhar mass, and mass
constraints from new MSPs will continue to map out the MSP mass
distribution. The PALFA pulsar PSR J1949+3106, which may also have a
higher mass than the Chandrasekhar mass \citep{dfc+12}, is an example
where future mass measurements may prove important.  Although high-DM
MSPs have timing precision problems associated with interstellar
scattering in addition to their being generally radio-faint, future
observations with next generation instruments like the Square
Kilometer Array (SKA) may be able to mitigate these factors.
Precision measurement of NS masses (e.g., Demorest et
al. 2010\nocite{dpr+10}) and the measurement of ultra-high spin rates
(e.g., Lattimer \& Prakash 2007; Hessels et al. 2006\nocite{lp07,
hrs+06}) can also rule out some high-density NS equations of state.

In this paper we report the discovery and follow-up timing of four
highly dispersed MSPs from the PALFA survey: PSRs J1844+0115,
J1850+0124, J1900+0308, and J1944+2236.  These MSPs all have DMs that
are in the top 5\% of non-PALFA field radio MSPs.  Three of the four
pulsars are in binary systems with low-mass companions.  These
discoveries bring the number of PALFA recycled pulsar discoveries to
15, including PSR J1903+0327 \citep{crl+08}, PSRs J1949+3106 and
J1955+2427 \citep{dfc+12}, and the partially recycled PSRs J2007+2722
and J1952+2630 \citep{kac+10,kla+11}. The six additional recycled
pulsars that have been confirmed in the survey need to be timed
further to establish their rotational and orbital characteristics.
These six pulsars will be published in forthcoming papers and are not
considered further here.  In \S2 we describe the discovery and
follow-up observations of the four pulsars, and we present their
phase-coherent timing solutions.  In \S3 we discuss our results, and
in \S4 we present our conclusions.

\section{Discovery, Timing, and Polarimetry  Observations}

\subsection{Discovery}

Until February 2009, survey data for the PALFA survey were recorded
with the Wideband Arecibo Pulsar Processor (WAPP) auto-correlation
spectrometers \citep{dsh00, cfl+06}.\footnote{Since 2009, the survey
has been recording data using the Jeff Mock spectrometers, which are
polyphase filterbanks that provide better radio frequency interference
rejection as well as increased bandwidth compared to the earlier
WAPP survey data.}  The WAPP backends had 3-level sampling and
provided 100 MHz of bandwidth for each of the 7 ALFA beams, centered
on a sky frequency of 1.4 GHz and split into 256 lags
(channels). These lags were sampled at 64 $\mu$s, and each survey
pointing was observed for 268 s. The four pulsars presented here were
discovered in data taken with this setup.

PALFA survey data are archived at the Cornell University Center for
Advanced Computing, where they are processed locally using custom
search software and are also staged for transport to other PALFA
consortium sites.  The data are processed at these sites with
dedicated computer clusters that use a search pipeline based on the
PRESTO software suite
\citep{r01,rem02}\footnote{http://www.cv.nrao.edu/$\sim$sransom/presto}.
Another analysis pipeline is also used which employs the huge
volunteer computing resources provided by the ``Einstein@Home''
project\footnote{http://einstein.phys.uwm.edu} to search the survey
data for binary systems with orbital periods as short as 11 minutes
\citep{kla+11}. Results from the PRESTO processing (including
candidate plots and associated information) are uploaded to a
collaborative web portal (www.cyberska.org), where interactive
applications are used to visually inspect and rate selected
candidates.

\begin{figure}
\centerline{\psfig{figure=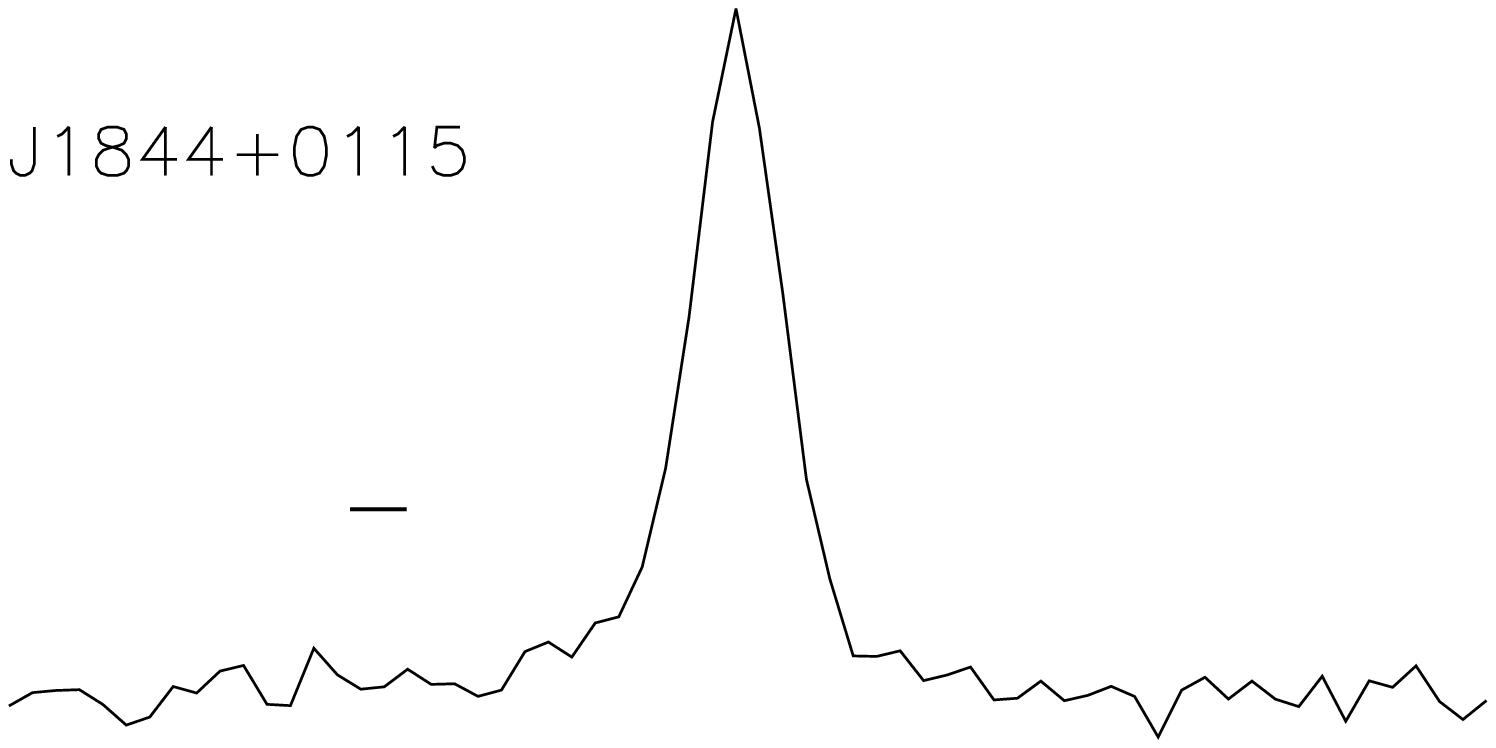,width=3.00in,angle=0}
\psfig{figure=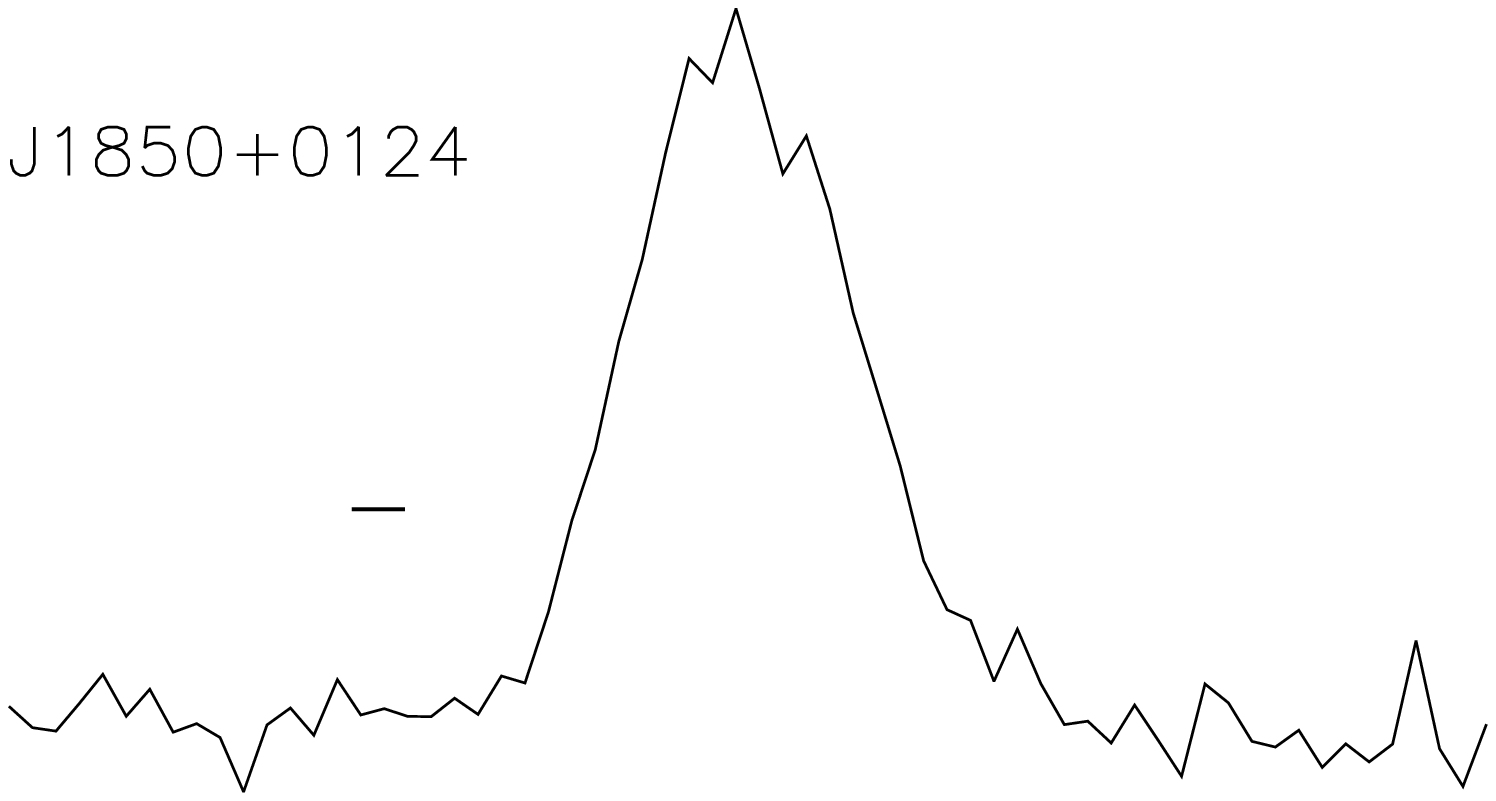,width=3.00in,angle=0}}
\centerline{\psfig{figure=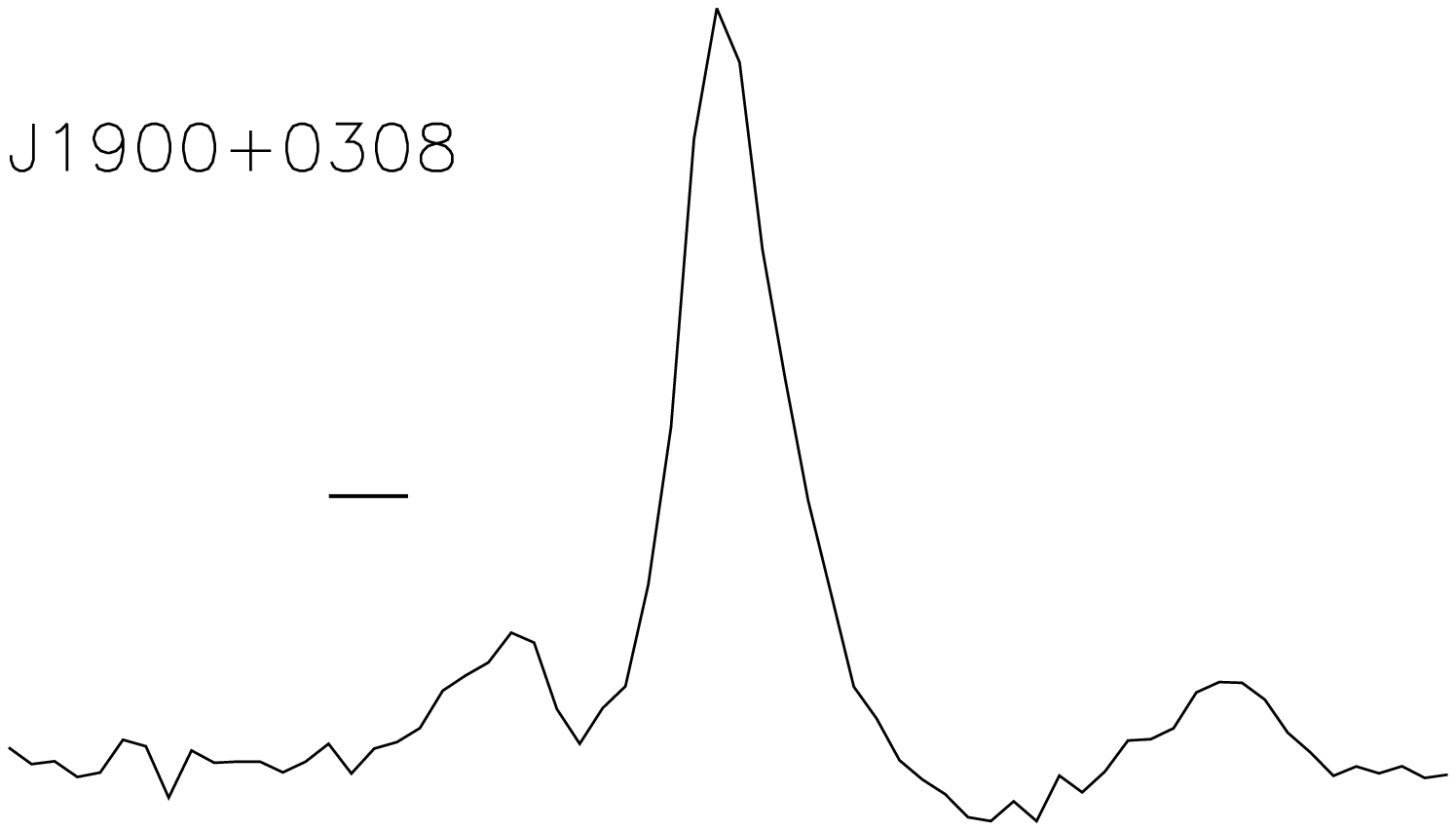,width=3.00in,angle=0}
\psfig{figure=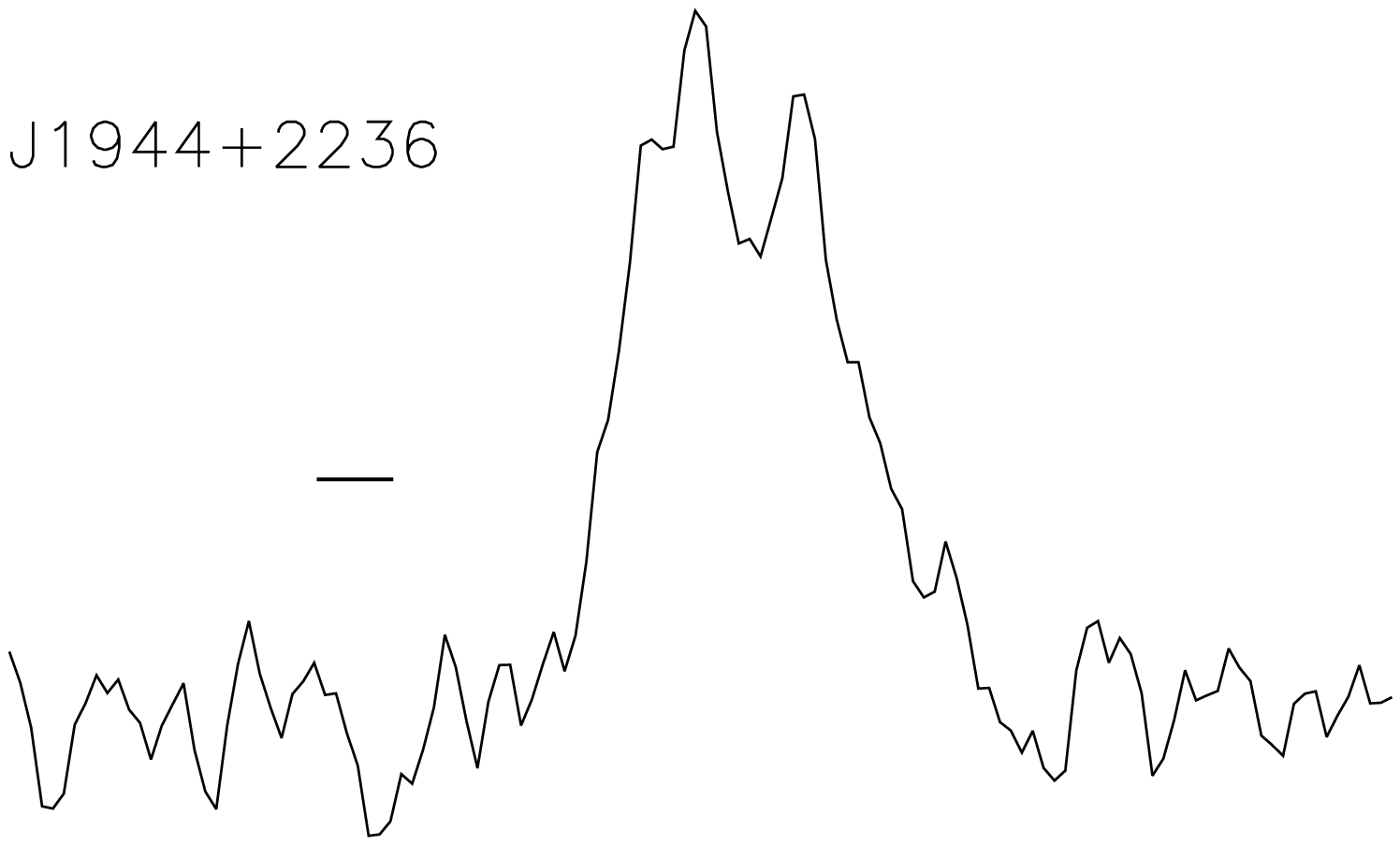,width=3.00in,angle=0}}
\caption{Integrated 1.4 GHz pulse profiles for four MSPs.  The pulse
profiles were constructed by phase aligning and adding between 30 and
50 minutes of Arecibo observations for each pulsar. A combination of
WAPP and Mock data were used for the profiles. In all cases, 64
profile bins were used, except for PSR J1944+2236 where 128 bins were
used.  One full period is shown in each case. The vertical axis is
arbitrary and has been scaled so that all profiles have the same
maximum value.  The horizontal bars indicate the amount of dispersion
smearing within channels for Arecibo observations taken with the
WAPPs.\label{fig-profiles}}
\end{figure}

The four MSPs presented here were all discovered with the PRESTO
search pipeline in survey pointings taken at different epochs (see
Table \ref{tbl-1} for the discovery observation MJDs).  Figure
\ref{fig-profiles} shows the integrated pulse profiles for the four
MSPs obtained by phase-aligning and adding the folded profiles from
Arecibo 1.4 GHz observations.  All of the integrated profiles in
Fig. \ref{fig-profiles} have moderate widths ($\sim 15$-30\% of the
pulse period).  In two cases (PSRs J1844+0115 and J1850+0124), the
profiles are clearly single-peaked. For PSR J1900+0308, there is some
indication of a weaker leading component (and possibly also a trailing
component).  PSR J1944+2236 may have a hint of a secondary peak close
to the center of the profile.  None of the profiles shows any clear
evidence of scattering, which is not surprising: according to the
NE2001 model of \citet{cl02}, the pulse scattering time for these
pulsars is expected to be small in all cases at 1.4 GHz ($\la 0.04$
ms). \citet{bcc+04} have noted that that in some cases pulse
scattering times can be underestimated by the NE2001 model by up to an
order of magnitude (see their Fig. 6). However, even if this were the
case here, the scattering time would still be $\la 10$\% of the pulse
period in all cases.  At the large DMs of these pulsars, any flux
variability would likely be from intrinsic mechanisms rather than
scintillation. We see no significant variability or other
intermittency (e.g., eclipsing effects) in the observations.

\subsection{Timing}

Soon after the discovery of these pulsars, follow-up timing
observations began with the Arecibo telescope and the Lovell telescope
at the Jodrell Bank observatory.  The Arecibo observations used
several systems: the ALFA receiver and single-pixel L-wide receivers,
with data recorded with either the WAPPs or the Mock spectrometer.

For the Jodrell Bank observations, dual-polarization cryogenic
receivers covered 384 MHz of bandwidth centered at 1520 MHz. The
bandwidth was split into 0.5 MHz channels in a digital filterbank, and
each channel was sampled to provide 1024 samples per pulsar
period. The effective sampling time for our four pulsars was therefore
a few $\mu$s and depended on the pulsar period. Each pulsar was
observed for between 2400 and 7200 s per timing session, depending on
the pulsar.  The sampled data in each channel were folded at the
topocentric pulsar period in each case (this was predicted by an
ephemeris), and the channels were subsequently dedispersed at the
ephemeris DM and summed.

For the Arecibo observations, the WAPP systems were used which covered
two adjacent 50 MHz bands. Each WAPP had 512 lags and a sampling time
of 64 $\mu$s.  The pulsars were typically observed for 600 s in each
observation. For PSR J1900+0308, some observations used the Mock
spectrometers (see, e.g., Deneva et al. 2012\nocite{dfc+12}).

The folded profiles from the Arecibo observations were phase-aligned
and summed to produce a high signal-to-noise pulse template for each
pulsar (see Fig. \ref{fig-profiles}).  Each observation was
dedispersed and folded with the latest timing ephemeris, and folded
profiles from each observation were cross-correlated in the Fourier
domain with the profile template to obtain times-of-arrival (TOAs)
with uncertainties.  A separate pulse template was produced from the
Jodrell Bank observations and this was used to similarly produce the
Jodrell Bank TOAs.  Table \ref{tbl-1} shows the number of TOAs
generated from Arecibo and Jodrell Bank with their typical rms values.

We fit the resulting TOAs using the TEMPO software
package\footnote{http://tempo.sourceforge.net. Note that we also
obtained consistent results with the TEMPO2 package \citep{hem06}.}
and standard pulsar timing procedures (see, e.g., Freire et
al. 2011\nocite{fbw+11} for details).  Initial phase-connections for
the pulsars were obtained from the Jodrell Bank observations alone,
and these were subsequently supplemented with Arecibo TOAs at a
variety of frequencies.  The Arecibo TOAs allowed us to constrain the
DM for each pulsar.  All three binaries have small values for the
product of orbital eccentricity and projected semi-major axis (see
Table \ref{tbl-1}), so we used the ELL1 binary orbital model of
\citet{lcw+01} for the timing solutions.

\begin{figure}
\centerline{\psfig{figure=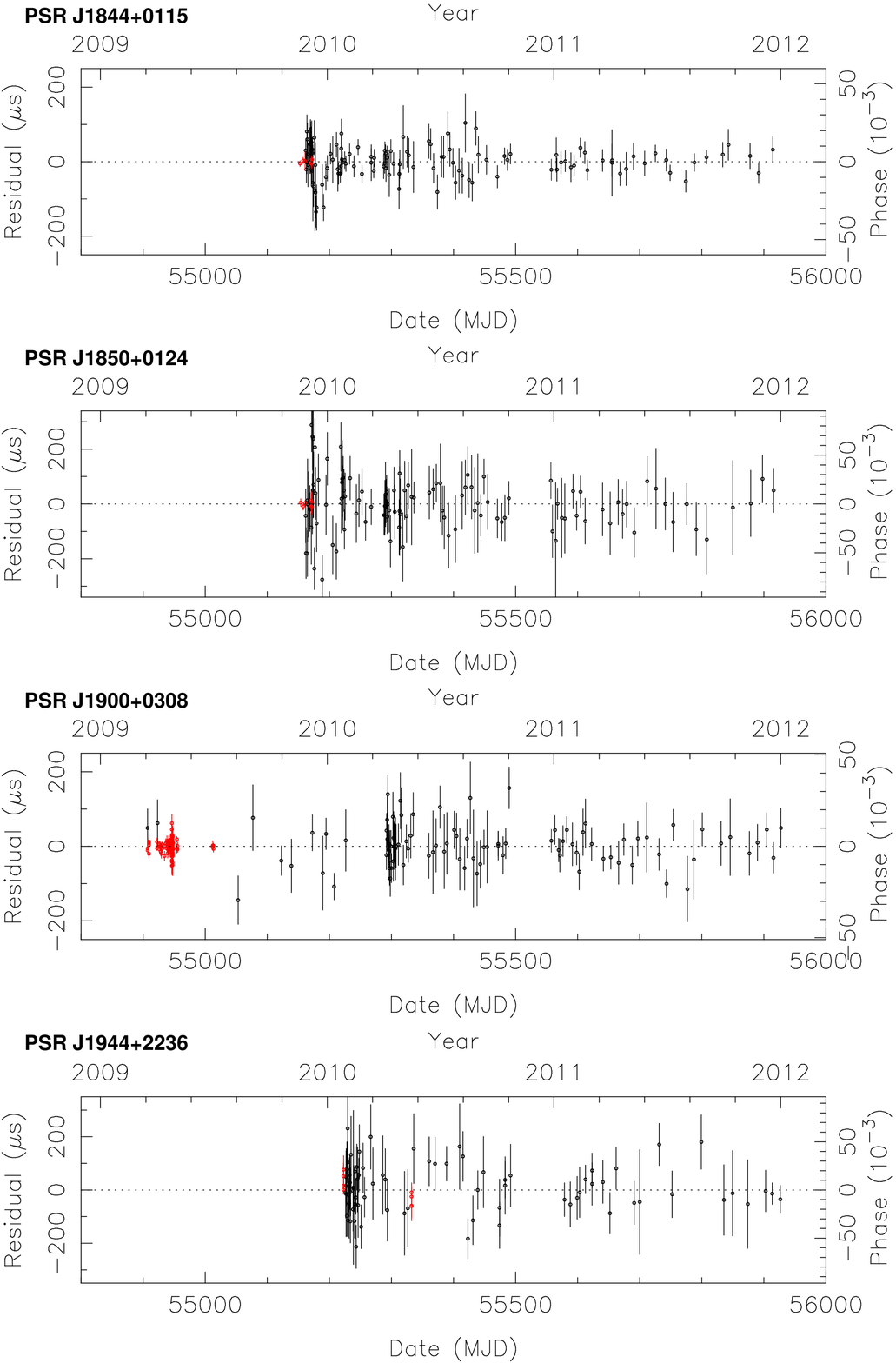,width=6in,angle=0}}
\caption{Timing residuals for four MSPs as a function of date.  Black
points indicate Jodrell Bank TOAs and red points indicate Arecibo
TOAs. The error bars shown were produced for each individual
TOA.\label{fig-1}}
\end{figure}

\begin{figure}
\centerline{\psfig{figure=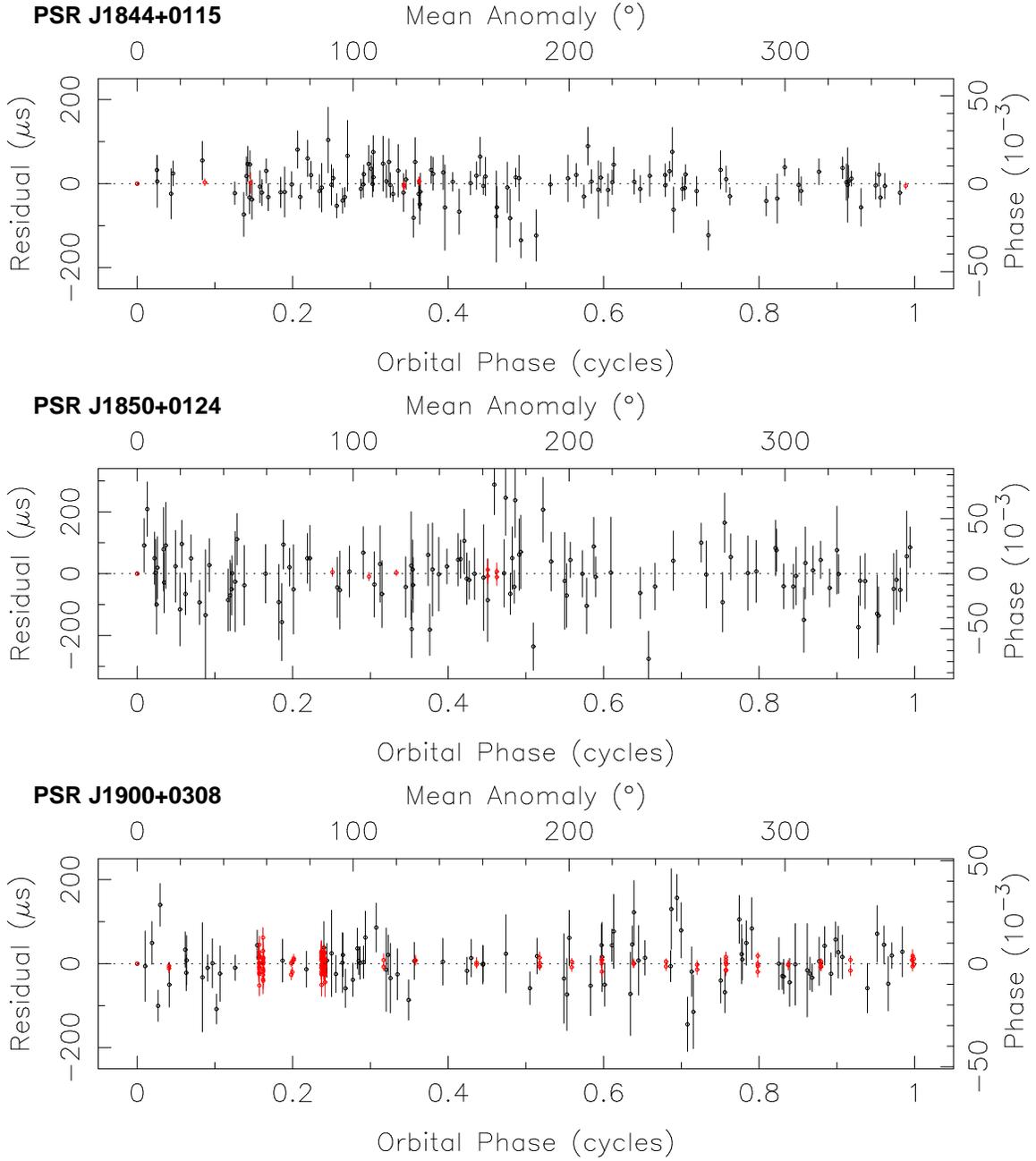,width=6in,angle=0}}
\caption{Timing residuals for three binary MSPs as a function of
orbital phase. PSR J1944+2236 is isolated and is not shown.  There are
no systematic trends in the residuals. None of the three binaries show
evidence of eclipsing effects which would be expected if the
companions were extended, non-degenerate stars with large orbital
inclination angles. This suggests that the companions are probably
low-mass white dwarfs, though more massive companions with very small
orbital inclinations cannot be ruled out.\label{fig-2}}
\end{figure}

Figures \ref{fig-1} and \ref{fig-2} show the timing residuals for the
four MSPs as a function of both date and orbital phase (except for PSR
J1944+2236, which is isolated). As commonly happens in MSP timing, we
found the formal TOA uncertainties in some data sets to be
underestimated by up to a factor of 2.0. We scaled the TOA
uncertainties by a common factor for each observational setup
(telescope and backend combination) in order to make the reduced
$\chi^2$ of each TOA subset equal to unity, and we used the scaled
uncertainties to weight the data in the timing fit. Increasing all
uncertainties by a factor until the normalized $\chi^2$ is one is a
conservative step, which produces less precise timing parameters, but
ones that are justified by the actual rms of the TOAs available. The
scaling factors are listed in Table \ref{tbl-1}. There are no obvious
systematic trends seen in the timing residuals, and the typical
residual rms values from the timing solutions are less than 2\% of the
pulse period in each case (Table \ref{tbl-1}).  These results indicate
that the model is adequately describing the TOAs, with no significant
unmodeled effects present.

The full timing solutions for the four MSPs are presented in Table
\ref{tbl-1}, with the listed uncertainties representing twice the
formal uncertainties produced by TEMPO. Apart from the directly
measured astrometric, spin, and orbital parameters, derived physical
parameters are also included. The physical parameters assume a pure
magnetic dipole spin-down. Estimated distances and luminosities for
the pulsars are also calculated and presented in the table. Note that
the Shklovskii effect \citep{s70} and accelerations from the Galactic
potential are not accounted for in the measured values of
$\dot{P}$. Until proper motions can be measured (see discussion
below), we cannot correct for this effect.

\subsection{Polarimetry}

All four pulsars were also observed in a set of 1.4 GHz Arecibo
polarimetry observations in order to measure polarization
characteristics in the pulse profiles and to obtain rotation measures
(RMs) and calibrated flux density estimates for the pulsars.  Each
pulsar was observed for between 5 and 15 minutes at a center frequency
of 1412 MHz using the ASP backend \citep{d07}. A total bandwidth of 24
MHz was split into 6 channels, each of 4 MHz, and the data were
coherently dedispersed and folded during the observation. The data
were then processed using PSRCHIVE tools \citep{hvm04}. The resulting
calibrated data files contained full Stokes parameters for each pulse
profile bin. The Stokes parameters were converted to total intensity,
linearly and circularly polarized intensity, and position angle (PA)
values.

\begin{figure}
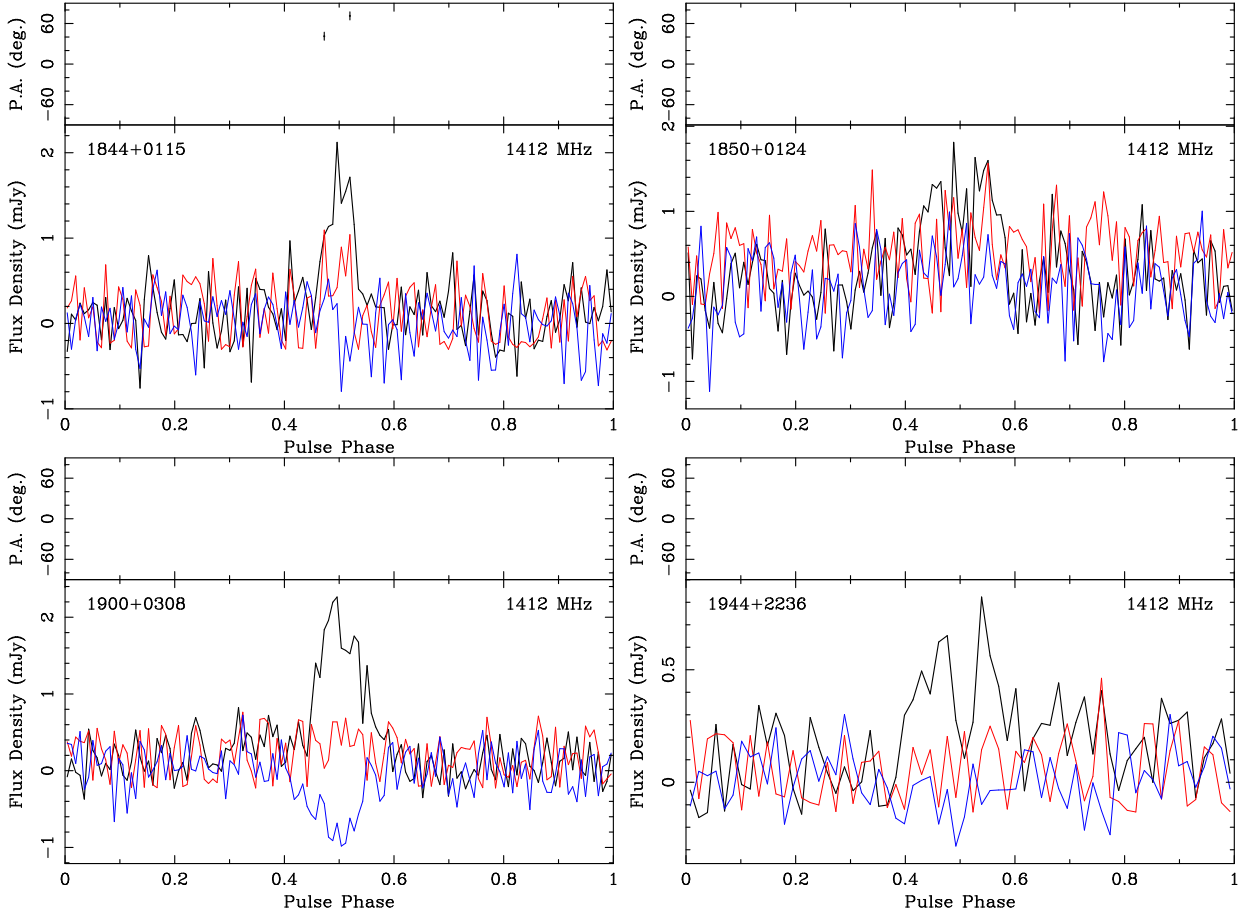

\centerline{
\psfig{figure=1844_poln.ps,width=3.2in,angle=270}
\psfig{figure=1850_poln.ps,width=3.2in,angle=270}}
\centerline{
\psfig{figure=1900_poln.ps,width=3.2in,angle=270}
\psfig{figure=1944_poln.ps,width=3.2in,angle=270}}
\caption{Polarization profiles for the four MSPs taken at a center
frequency of 1412 MHz with Arecibo using the ASP backend
\citep{d07}. RMs for the pulsars could not be determined from these
observations, so none of the profiles have been corrected for Faraday
rotation. Each plot has 128 phase bins, except PSR J1944+2236 which
has 64 bins. In the bottom part of each plot, the black, red, and blue
lines correspond to total, linearly polarized, and circularly
polarized intensity, respectively. The top part of each plot shows the
linear polarization PA for bins with S/N $> 3$. The low signal
strength in the observations is likely to be responsible for the lack
of any significant linear polarization observed and any measurable PAs
across the profiles.\label{fig-poln}}
\end{figure}

We attempted to search for the RM for each pulsar in order to correct
the data for Faraday rotation. We used trial RM values to see which
trial produced the maximum linear polarization when the frequency
channels were summed.  This would indicate that the correct RM had
been found.  We were unable to find a reliable RM estimate in any of
the four cases. Figure \ref{fig-poln} shows the polarization profiles
for the four pulsars without any Faraday rotation correction. None of
the pulsars shows clearly significant polarization, with the possible
exception of PSR J1900+0308, which may have an excess of right-handed
circular polarization across the on-pulse bins. None of the pulsars
has a linearly polarized intensity that is significant enough for
reliable PA measurements across the on-pulse bins (in fact, only PSR
J1844+0115 has any measurable PAs at all).  Faraday smearing across
the band is unlikely to be a significant factor in the small measured
polarized signal. For RMs as high as a few hundred rad m$^{-2}$, the
uncorrected Faraday rotation would reduce the linear polarization by
only a few percent.

The calibrated files were also used to determine 1.4 GHz flux
densities for the pulsars (Table \ref{tbl-1}). In all cases except PSR
J1944+2236 we obtained flux density estimates, but only one digit of
precision is quoted owing to the significant uncertainties in these
estimates. For PSR J1944+2236, only an upper limit of 0.1 mJy is
quoted for the flux density since a value was not measured. This was
chosen as a reasonable upper limit since this is the value measured
for the next weakest pulsar, and if PSR J1944+2236 had this flux
density it would have likely been measurable.  Longer observations in
the future using the ASP or different observations using a system with
a wider bandwidth may be useful in obtaining RM measurements and a
clearer indication of the polarization characteristics of these MSPs.

\section{Discussion}

The inferred spin-down luminosities, surface magnetic field strengths,
and characteristic ages of these four pulsars (Table \ref{tbl-1}) are
typical of the values of most fully recycled pulsars.  Likewise, the
orbital properties of the three binaries presented here are typical,
and they have low-mass companions ($M_{c} \sim 0.2 M_{\odot}$) and low
eccentricities ($e \la 3 \times 10^{-4}$ in all cases), which suggests
that the systems are fully recycled with white dwarf companions.

The measured eccentricities of the three binaries can be compared to
the relationship between the binary orbital period, $P_{b}$, and
eccentricity, $e$, that was outlined by \citet{p92} and \citet{pk94}
for stable mass transfer from a Roche-lobe filling red giant (see also
Camilo et al. 2001\nocite{clm+01} and Lorimer 2008\nocite{l08}). PSRs
J1844+0115 and J1850+0124 have eccentricities that easily fall within
the predicted range of the model for their orbital periods.  PSR
J1900+0308 has an eccentricity that is small compared to the predicted
range but is still consistent with the model.

PSRs J1850+0124 and J1900+0308 may be useful for tests of the Strong
Equivalence Principle (SEP) and similar deviations from general
relativity \citep{ds91,bd96,sfl+05,gsf+11}.  The figure of merit for
SEP tests is $P_{b}^{2}/e$, which for PSR J1850+0124 is large, $\ga
10^8$ day$^{2}$. This places it in the same range as PSRs
J1711$-$4322, J1933$-$6211, and J1853+1303 (see Table 1 of
\citet{gsf+11}), which were among the best pulsars used to constrain
violation of the SEP and the strong-field version of the Parameterized
Post-Newtonian (PPN) parameter $\alpha_3$ in that paper
\citep{wn72,de92}. 
 
PSR~J1900+0308 has a very large lower limit for $P_{b}^{1/3}/e$,
making it potentially a key pulsar for constraining the (strong-field)
PPN parameter $\hat{\alpha_1}$ \citep{de92a,bcd96}.  Historically, the
pulsars used for this test have had orbital periods of just a few
days, since these tend also to have the lowest orbital eccentricities.
Unfortunately, the short orbital periods of the relevant pulsars make
it hard to define a population of systems thought to have followed the
same general evolutionary path; in particular, pulsars with
white-dwarf companions and orbital periods under about 4 days cannot
safely be assumed to follow the $P_{b}$--$m_{2}$ relation
\citep{rpj+95,ts99,tc99}.  Since a well-defined population for which
one can make reasonable evolution-based guesses at the pulsar and
companion masses is desirable to mitigate against selection effects
\citep{wex00}, it has been difficult to define a suitable set of
pulsars for use in the $\hat{\alpha_1}$ test.  The existence of
PSR~J1900+0308, assuming it does follow the $P_{b}$--$m_{2}$ relation,
opens the possibility of using this class of pulsar for the
$\hat{\alpha_1}$ test.  This will best be done once the proper motion
and eccentricity are well-measured for this system.

We are not currently able to measure proper motions for these
pulsars. Making the assumption of a typical transverse speed of $\sim
100$ km s$^{-1}$ and a representative distance of 5 kpc, the proper
motion would be $\sim 4.2$ mas yr$^{-1}$. This would introduce a
timing residual of $\sim 5$ $\mu$s after one year \citep{s70}, and the
timing spans for these pulsars are relatively short ($\sim 2$ yr for
three of the four pulsars; see Table \ref{tbl-1}).  The relatively
large rms values of the residuals (of order tens of $\mu$s) preclude
these objects from being useful in pulsar timing arrays in the search
for gravitational waves.  Given their large distances, their proper
motions are likely to be small and difficult to measure given their
timing precision, making it hard to accurately determine the kinematic
contributions to $\dot{P}$ and $\dot{P_{b}}$.  

We performed a test with the timing residuals in order to assess the
role of red noise (if any) in the timing behavior of the four
pulsars. The test we employed was the zero-crossing test, which
measured the number of times the residuals change sign (cross zero) as
a function of time (e.g., Deneva et al. 2012\nocite{dfc+12}). The
validity of this test is independent of the spacing of the individual
timing points and corresponding residuals. The expected number of zero
crossings of $N$ residuals if only white noise were present is
$\langle Z_{W} \rangle = (N-1)/2$, with an uncertainty in the mean of
$\sigma_{Z_{W}} = \sqrt{(N-1)}/2$. The presence of timing noise (red
noise) will produce fewer crossings than expected. In all four cases,
the observed number of residual crossings fell within the expected
range $\langle Z_{W} \rangle \pm \sigma_{Z_{W}}$, indicating that red
noise is not significant in the timing behavior of these pulsars at
the current (rather poor) level of precision.  We have also
characterized the timing noise in these 4 MSPs using a relative timing
noise parameter, $\zeta$, outlined by \citet{sc10}.  $\zeta$ is the
ratio of the observed timing residual value to a value expected from
only timing noise. This latter parameter is determined using a scaling
law (see their Eq. 7). The parameters used in the scaling were derived
from measurements of canonical and millisecond pulsars (see Table 1 of
Shannon \& Cordes 2010). For all four MSPs, we found that this ratio
was $\zeta \gg 1$. This indicates that white noise is dominating the
residuals in all cases, and it is consistent with the results of the
zero-crossing test above.

The relatively large distances to these pulsars are restrictive in
terms of their potential for multi-wavelength follow-up, and,
unsurprisingly, the SIMBAD
database\footnote{http://simbad.u-strasbg.fr/simbad/} indicates that
there are no optical or IR counterparts listed at the positions of two
of the three binary pulsars.  There is an unidentified IR source, IRAS
18421+0112, located $\sim 22''$ from PSR J1844+0115 which has a
position uncertainty of $17''$ along the major axis of its position
ellipse, but there is nothing listed in the {\it 2MASS} catalog at
this location.  It seems unlikely that this IRAS source is the binary
counterpart of PSR J1844+0115, since the pulsar shows no evidence of
eclipses or other variability/intermittency in the timing residuals
that we might expect from a non-degenerate, extended companion (see
Fig. \ref{fig-2}), and we see none of the timing jitter that is
clearly present in the two known cases of these kinds of Galactic MSP
systems that have been published to date (PSR J1023+0038, Archibald et
al.  2009\nocite{asr+09}; and PSR J1723$-$2837, Crawford et
al. 2010\nocite{clm+10}).

There are also no known X-ray or $\gamma$-ray counterparts in the
HEASARC catalog at the four pulsar positions. This is not surprising
since $\dot{E} \la 10^{34}$ erg s$^{-1}$ and for this spin-down
luminosity range, the X-ray emission would be quite faint.  To search
for $\gamma$-ray pulsations from the four MSPs, we retrieved {\it
Fermi} LAT data from the start of the mission up to 2012 February 6
using $1^{\circ}$ extraction radii centered on the pulsar
positions. The event lists were filtered using the recommended cuts in
maximum zenith angle ($100^{\circ}$), event class (2), and photon
energy ($> 100$ MeV). The $\gamma$-ray events were folded using the
Fermi plug-in for TEMPO2\footnote{See
http://fermi.gsfc.nasa.gov/ssc/data/analysis/user/Fermi\_plug\_doc.pdf.}
and the radio ephemerides from Table 1. In all instances, no
statistically significant pulsations were detected.  Repeating the
analysis with a $>$300 MeV energy cut and extraction radii in the
range $0.5^{\circ}-1^{\circ}$ also yielded no detections.

\begin{figure}
\centerline{\psfig{figure=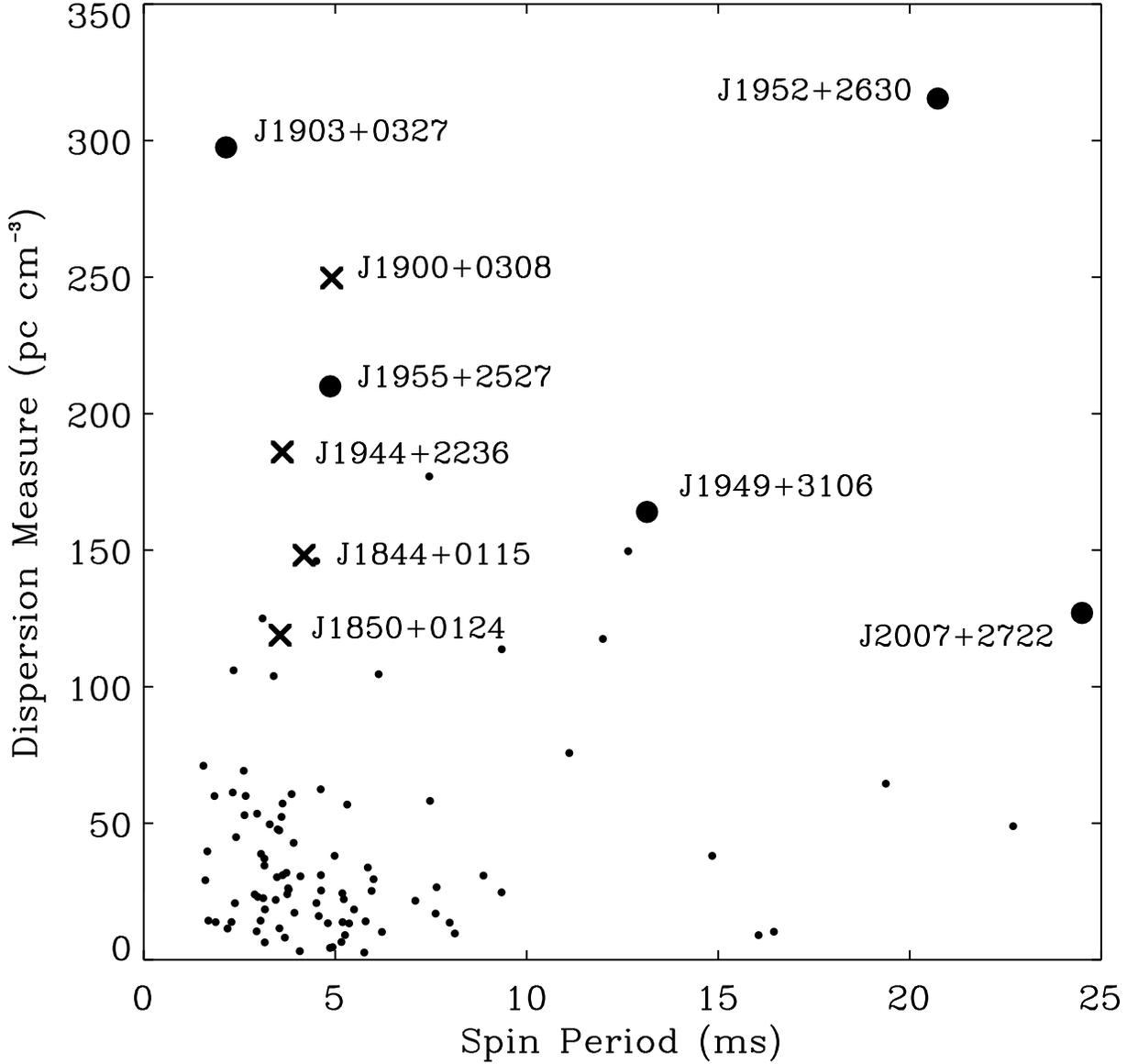,width=7in,angle=0}}
\caption{Dispersion measure vs. spin period for 90 Galactic field
radio MSPs from the ATNF pulsar catalog (only pulsars with periods
less than 25 ms and $\dot{P} < 10^{-17}$ are plotted, and GC and
radio-quiet pulsars are not included). These pulsars are plotted as
dots.  Also plotted are the four PALFA MSPs reported in this paper
(crosses) and five other MSPs discovered by PALFA (filled circles)
\citep{crl+08, kac+10, kla+11, dfc+12}. All nine PALFA pulsars are
labeled. It is clear that the PALFA survey is exploring new parameter
space with the discovery of these pulsars. The DMs of these pulsars
are among the highest that have been previously discovered for MSPs in
the Galaxy.\label{fig-5a}}
\end{figure}

All four of these MSPs share the quality that they have very large DMs
and DM-inferred distances. This is also a feature of the other MSPs
that have been discovered by the PALFA survey.  Previous large-scale
surveys had poorer sensitivity to these kinds of MSPs owing to their
lower observing frequencies and inadequate observing instrumentation
(wider frequency channels and insufficient sampling rates).  Figure
\ref{fig-5a} illustrates this with a plot of DM vs. spin period for
all 90 non-PALFA Galactic field radio pulsars in the ATNF pulsar
catalog having $P < 25$ ms and $\dot{P} < 10^{-17}$
\citep{mht+05}\footnote{http://www.atnf.csiro.au/research/pulsar/psrcat/}.
This group of recycled pulsars does not include globular cluster (GC),
radio quiet, or young pulsars. Also shown are the four MSPs described
here and the five other PALFA MSPs that have been reported to date
\citep{crl+08, kac+10, kla+11, dfc+12}.  It is clear that the PALFA
MSPs occupy an area of the DM-period phase-space where very few MSPs
are currently known.

\begin{figure}
\centerline{\psfig{figure=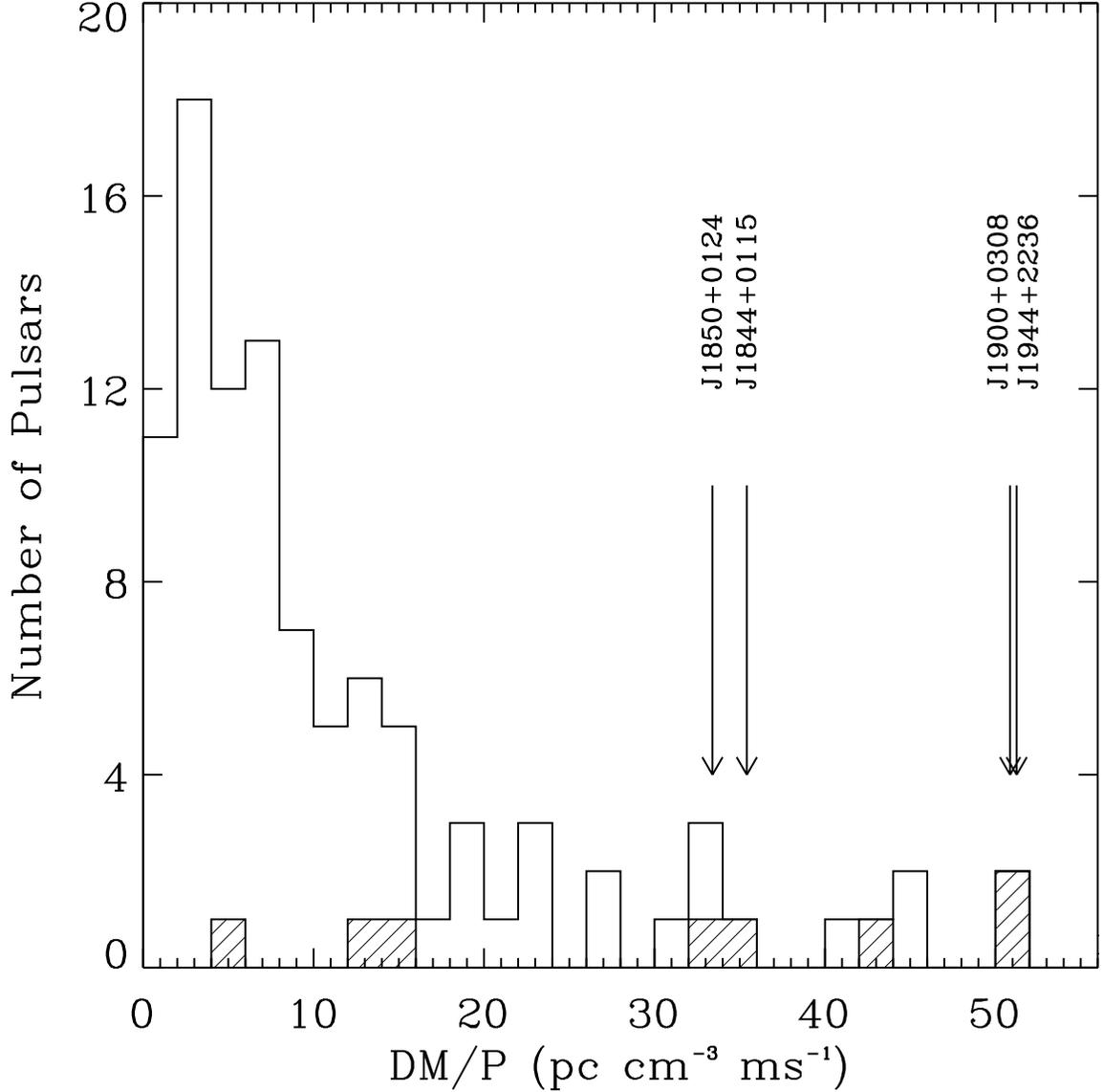,width=7in,angle=0}} 
\caption{Histogram of DM/$P$ for 90 known field radio pulsars from the
ATNF catalog and 8 PALFA MSPs (not including PSR J1903+0327) having $P
< 25$ ms, $\dot{P} < 10^{-17}$. GC and radio-quiet pulsars are not
included here.  PSR J1903+0327 is not shown owing to its very large
DM/$P$ value of 138.4 pc cm$^{-3}$ ms$^{-1}$, which is far off the
scale. The unshaded histogram shows all 98 pulsars, including the 8
PALFA pulsars. The shaded histogram shows the subset of 8 PALFA MSPs.
The four MSPs presented in this paper are indicated by arrows and are
labeled. It is clear that the PALFA survey is finding fast, distant
pulsars that were undetectable in previous large-scale
surveys.\label{fig-pdm}}
\end{figure}

\begin{figure}
\centerline{\psfig{figure=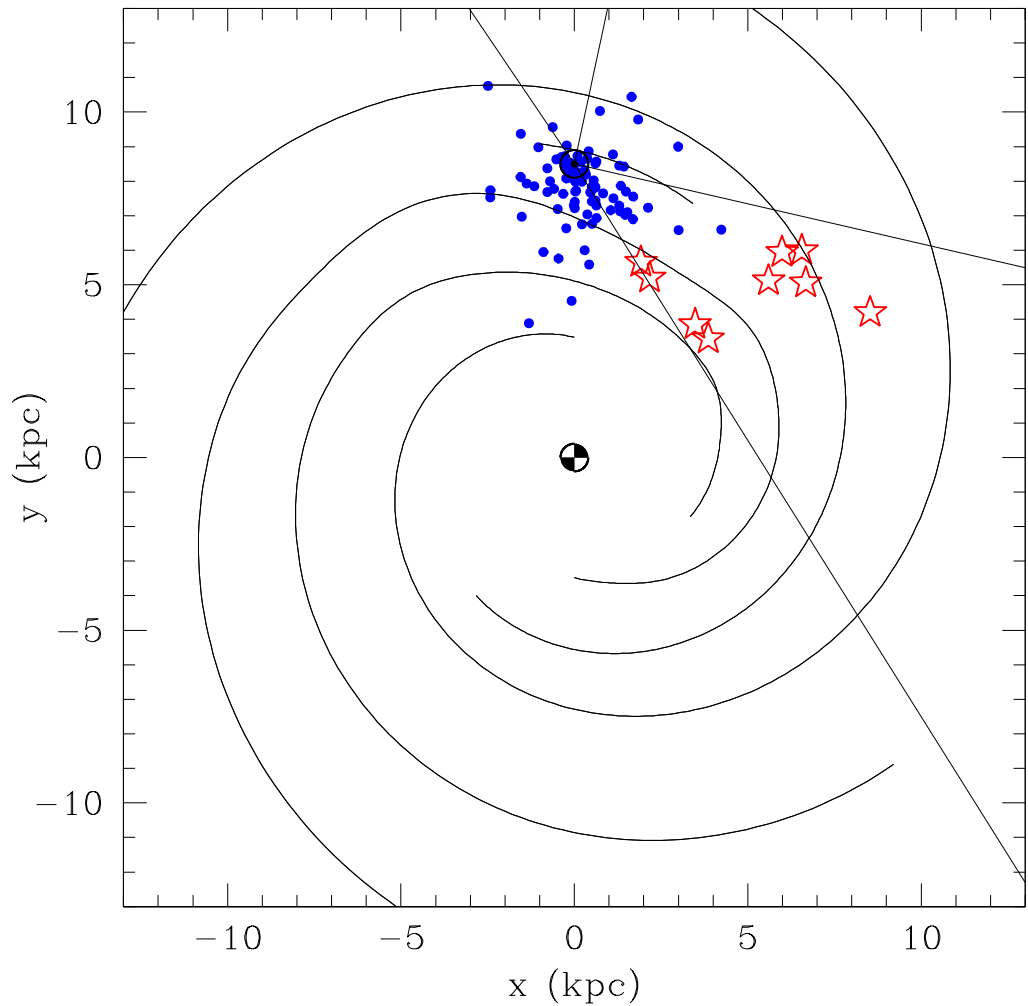,width=6.0in,angle=0}} 
\caption{Overhead Galactic projection plot showing the nominal
locations of 90 known field radio MSPs in the ATNF catalog with $P <
25$ ms and $\dot{P} < 10^{-17}$ (blue dots) and 9 known PALFA MSPs
(red stars) \citep{crl+08,kac+10,kla+11,dfc+12}, including the 4
described in this paper.  The nominal distances are inferred from the
DMs and are uncertain by $\sim 30$\%. The Sun is located at the center
of the cluster of dots, and the Galactic center is at the origin. The
black curves indicate locations of the Galactic spiral arms according
to the NE2001 model of Cordes \& Lazio (2002). The cones indicated by
the two pairs of straight lines indicate the regions of the Galactic
plane that are visible from Arecibo. The PALFA MSPs are among the most
distant non-cluster MSPs known.\label{fig-gal}}
\end{figure}

Figures \ref{fig-pdm} and \ref{fig-gal} also illustrate the power of
the PALFA survey for finding fast, distant MSPs. Both figures include
the same set of data as shown in Fig. \ref{fig-5a}. Fig. \ref{fig-pdm}
shows a histogram of the ratio DM/$P$ (excluding PSR J1903+0327, since
it falls well beyond the plot limits). The subset of PALFA MSPs are
indicated with shaded parts of the histogram.  The positions of the 4
MSPs presented here are indicated with arrows, all of which fall near
the edge of the histogram.  This plot demonstrates how we are
exploring a larger DM/$P$ parameter space with this
survey. Fig. \ref{fig-gal} shows a Galactic projection plot of the
same set of pulsars. The nominal locations of the PALFA MSPs in the
Galactic plane are indicated by stars.  Exploring the high DM/$P$
parameter space is necessary for a full Galactic census of MSPs.

To demonstrate the potential of the PALFA survey in such a population
analysis, we have carried out a preliminary investigation to model
DM/$P$ for Galactic MSPs. We used the freely available PSRPOP software
package\footnote{http://psrpop.sourceforge.net} to carry out this
work, and we generated simple ``snapshot'' models of the population
which are normalized to reproduce the yield of 20 MSPs detected in the
PMPS (e.g. Lorimer et al.~2006\nocite{lfl+06}).  The simulation
procedure is described in that paper and also in \citet{sks+09}. In
brief, we generated four different models (labeled A through D) of the
present day MSP population which matched the number of MSPs detected
by the PMPS, but produced a variety of different DM/$P$ values and
predicted yields for the PALFA survey. For the purposes of this work,
we focused on changing assumptions about the spatial distribution of
the underlying MSP population.

Model A assumes a Gaussian radial density profile with a standard
deviation $\sigma_R= 6.5$~kpc, an exponential scale height with a mean
of $h_z = 500$~pc, the log-normal 1400-MHz pulsar luminosity function
found by \citet{fk06}, and the MSP period distribution and beaming
model used in \citet{sks+09}. Model B is similar to Model A with the
exception that the radial density is uniform (i.e., a constant surface
density) throughout a disk of radius 25~kpc. Models C and D both
follow Model A, but with $h_z$ set to 250 and 750 pc, respectively.
Further models should be explored in a subsequent analysis, but for
the purposes of the present work, the above choices demonstrate the
importance of DM/$P$ as a diagnostic for population studies.

The results of our simulations are shown in Table \ref{tbl-dunctable}
where we tabulate, along with the observed sample, the numbers of
detectable MSPs along with the median DM/$P$ found in each
simulation. By definition, due to the way we normalize the
simulations, the number of detected MSPs in the PMPS, $N_{\rm PMPS}$,
is 20. For models A through D, we tabulated the predicted number of
MSPs detectable in the PALFA survey, $N_{\rm PALFA}$.  The median
DM/$P$ values for each simulation can be compared with those currently
observed in the survey samples. As can be seen, there is a tremendous
variation in the predicted values, and the median DM/$P$ depends
strongly on the choice of scale height and radial distribution.  Model
B appears to underestimate both the observed number of PALFA MSPs
(assuming we find more as the survey proceeds) as well as the median
DM/$P$ value. Model C seems to overestimate the median DM/$P$
value. Models A and D are most consistent with our observations to
date, suggesting that a large scale height $h_z$ and a Gaussian radial
profile density are favored.

\section{Conclusions} 

With its high time and frequency resolution and relatively high observing
frequency, the PALFA survey is sensitive to Galactic field MSPs at
large distances and DMs.  The survey has so far discovered a total of 15 MSPs
with $P < 25$ ms, 14 of which have DM $> 100$ pc cm$^{-3}$. Four of
these PALFA MSPs (PSRs J1844+0115, J1850+0124, J1900+0308, and
J1944+2236) are presented here with phase-coherent timing solutions.
As an ensemble, the PALFA MSP discoveries show the ability of PALFA to
extend the volume of MSP discovery space to relatively high DMs.  This
is the first step toward a nearly complete census of the Galactic MSP
population, which will be possible with the SKA.

\acknowledgements

The Arecibo observatory is operated by SRI International under
cooperative agreement with the National Science Foundation
(AST-1100968) and in alliance with Ana G. Mendez-Universidad
Metropolitana, and the Universities Space Research Association.  This
work was also supported by NSF Grant AST-0807151, by NSERC Discovery
Grants, by FQRNT via the Centre de Recherche Astrophysique du Quebec,
by CIFAR, by CANARIE, by Compute Canada, by the Canada Foundation for
Innovation, and a Killam Research Fellowship.  VMK holds the Lorne
Trottier Chair in Astrophysics and Cosmology and a Canadian Research
Chair in Observational Astrophysics.  JWTH acknowledges funding from
an NWO Veni Fellowship. BK gratefully acknowledges the support of the
Max Planck Society. PL was partly funded by a NSERC PGS scholarship
and an IMPRS fellowship. PF gratefully acknowledges the financial
support by the European Research Council for the ERC Starting Grant
BEACON under contract no. 279702.

\clearpage

\begin{deluxetable}{lcccc}
\tabletypesize{\tiny}
\tablecaption{Timing Parameters for Four MSPs\label{tbl-1}}
\tablewidth{0pt}
\tablehead{
\colhead{PSR} &
\colhead{J1844+0115} &
\colhead{J1850+0124} &
\colhead{J1900+0308} &
\colhead{J1944+2236} 
}
\startdata
Right ascension (J2000) &  18:44:40.5474(3)          &  18:50:01.0139(8)  &  19:00:50.5548(3)   &  19:44:01.0707(11) \\      
Declination (J2000)     & +01:15:34.974(11)          & +01:24:34.61(3)    & +03:08:24.079(13)   & +22:36:22.62(2) \\ 	         
Spin frequency, $f$ ($s^{-1}$)          & 238.91757323111(15)        & 280.9175173300(4)  & 203.6975581461(2) & 276.3963275801(5) \\     
Frequency derivative, $\dot{f}$ ($s^{-2}$)    & $-6.12(12) \times 10^{-16}$ & $-8.6(4) \times 10^{-16}$   & $-2.45(9) \times 10^{-16}$  & $-5.7(4) \times 10^{-16}$  \\ 
Dispersion measure, DM (pc cm$^{-3}$) & 148.22(2)    & 118.89(5) 	  & 249.898(11)           & 185.45(12) \\	    	 
Orbital period, $P_{b}$ (d)  & 50.6458881(11)  	     & 84.949858(4) 	  & 12.47602144(10)       & $-$ \\	
Time of periastron passage, $T_{0}$ (MJD)\tablenotemark{a} & 55409.21(4)  	          & 55241.3(3)          & $-$     & $-$ \\	      
Projected semi-major axis, $x$ (s)\tablenotemark{b}& 14.173495(9) 	  & 34.00102(2)       & 6.716377(4)         & $-$ \\		       
Longitude of periastron, $\omega$ (deg)\tablenotemark{a} & 189.6(3) & 98.8(11)            & $-$               & $-$ \\		
Eccentricity, $e$       & 2.578(11) $\times 10^{-4}$ & 6.90(12) $\times 10^{-5}$  & $< 3.3 \times 10^{-6}$          & $-$ \\      
Time of ascending node, $T_{asc}$ (MJD)\tablenotemark{c} &  55382.532417(5) & 55217.963824(11) & 55304.460562(3) & $-$ \\ 
$\epsilon_{1} = e \sin \omega$\tablenotemark{c} & $-4.32(13) \times 10^{-5}$  & $6.82(12) \times 10^{-5}$  & $< 3.0 \times 10^{-6}$ & $-$ \\ 
$\epsilon_{2} = e \cos \omega$\tablenotemark{c} & $-2.542(11) \times 10^{-4}$ & $-1.05(13) \times 10^{-5}$ & $< 3.1 \times 10^{-6}$ & $-$ \\ 
                       &                             &                    &                     &     \\ 
Period, $P$ (ms)        & 4.185543936664(3) & 3.559763768043(5) & 4.909239016417(4) & 3.617993078111(6) \\  
Period derivative, $\dot{P}$   & $1.07(2) \times 10^{-20}$  & $1.09(5) \times 10^{-20}$  & $5.9(2) \times 10^{-21}$ & $7.5(5) \times 10^{-21}$ \\ 
Mass function ($M_{\odot}$) &  0.001191860(2)        & 0.005848349(13)    & 0.002089949(4)      & $-$ \\
Companion mass ($M_{\odot}$)\tablenotemark{d} & $> 0.14$ & $> 0.25$ & $> 0.17$                  & $-$ \\ 
Galactic longitude, $l$ (deg) & 33.28                & 34.02              & 36.79               & 58.90 \\
Galactic latitude, $b$ (deg)  & +2.08                & +0.96              & $-$0.66             & $-$0.66 \\ 
Surface magnetic field, $B$ (G)\tablenotemark{e} & $2.1 \times 10^{8}$    & $2.0 \times 10^{8}$     & $1.7 \times 10^{8}$  & $1.7 \times 10^{8}$ \\
Spin-down luminosity, $\dot{E}$ (erg s$^{-1}$)\tablenotemark{e} & $5.8 \times 10^{33}$  & $9.6 \times 10^{33}$  & $2.0 \times 10^{33}$ & $6.4 \times 10^{33}$ \\
Characteristic age, $\tau_{c}$  (Gyr)\tablenotemark{e} & 6.2 & 5.2        & 13.0                & 7.5 \\
Distance, $d$ (kpc)\tablenotemark{f} & 3.9 & 3.4 & 5.8 & 6.5 \\
Distance from Galactic plane, $|z|$, (kpc)\tablenotemark{g} & 0.14 & 0.06 & 0.07 & 0.08 \\
1400 MHz flux density, $S_{1400}$ (mJy) & $\sim 0.1$ & $\sim 0.2$         & $\sim 0.2$          & $\la 0.1$     \\
1400 MHz radio luminosity, $L_{1400}$ (mJy kpc$^{2}$)\tablenotemark{h}  & $\sim 1.5$ & $\sim 2.3$  & $\sim 6.7$  & $\la 4.2$ \\
                       &                             &                    &                     &     \\ 
Discovery observation MJD & 53493                    & 54555              & 53656               & 53647       \\ 
Timing epoch (MJD)     & 55383   	             & 55383              & 55255               & 55415       \\	     	   	
TOA range (MJD)        & 55152-55914                 & 55153-55915        & 54907-55927         & 55222-55926 \\ 
Timing span (d)        & 762                         & 762                & 1020                & 704         \\
Number of points in timing fit (Jodrell/Arecibo) & 116/7  & 110/7         & 94/136              & 77/8        \\
Characteristic TOA residual ($\mu$s) (Jodrell/Arecibo) & 40/10  & 100/10    & 80/10               & 80/10       \\ 
TOA error scale factor (Jodrell/WAPP/Mock)\tablenotemark{i} & 1.20/1.00/$-$ & 2.03/1.00/$-$ & 1.09/1.25/1.45 & 1.87/$-$/1.85  \\
Weighted rms post-fit residual  ($\mu$s) & 25.4      & 51.6               & 15.8                & 66.1 
\enddata

\tablecomments{Figures in parentheses are uncertainties in the last
digit quoted and are twice the formal errors from the TEMPO timing
solution.}

\tablenotetext{a}{The parameters $T_{0}$ and $\omega$ are highly
covariant in the timing solutions. Observers should use the following
values of $T_{0}$ and $\omega$, respectively: 55409.21301273 and
189.6504286222 for PSR J1844+0115; 55241.26871848 and 98.7613423916
for PSR J1850+0124.}

\tablenotetext{b}{$x = a \sin i / c $ where $a$ is the semi-major axis and
$i$ is the orbital inclination angle.}

\tablenotetext{c}{For all three binaries, the extremely small
eccentricity ($e \ll 1$) introduces a covariance term between the time
of periastron passage $T_{0}$ and the longitude of periastron
$\omega$. In these cases the ELL1 binary model was used
\citep{lcw+01}, where the time of ascending node $T_{asc}$, defined as
when $\omega = 0$, and $\epsilon_{1} = e \sin \omega$ and
$\epsilon_{2} = e \cos \omega$ are fit instead. $T_{0}$, $e$, and
$\omega$ are derived from this.}

\tablenotetext{d}{Assumes an inclination angle $i = 90^{\circ}$ and a
pulsar mass of 1.35 solar masses.}

\tablenotetext{e}{$B = 3.2 \times 10^{19} (P \dot{P})^{1/2}$; $\dot{E}
= 4 \pi^{2} I \dot{P} / P^{3}$, with an assumed moment of inertia $I =
10^{45}$ g cm$^{2}$; $\tau_{c} = P / 2 \dot{P}$. Note that these
parameters depend on $\dot{P}$ which may be affected by the Shklovskii 
effect.}



\tablenotetext{f}{From the NE2001 DM-distance model of \citet{cl02}.}

\tablenotetext{g}{$|z| = d \sin |b|$} 

\tablenotetext{h}{$L_{1400} = S_{1400} d^{2}$}

\tablenotetext{i}{TOA uncertainties from each instrumental setup were
multiplied by this factor (EFAC in TEMPO) to correct for the generally
underestimated uncertainties produced by TEMPO (see the text for
justification).}

\end{deluxetable}

\begin{deluxetable}{lcccc}
\tablecaption{MSP Population Simulation Results for Four Models\label{tbl-dunctable}}
\tablewidth{0pt}
\tablehead{
\colhead{Sample} &
\colhead{$N_{\rm PMPS}$\tablenotemark{a}} &
\colhead{$N_{\rm PALFA}$\tablenotemark{a}} &
\colhead{DM/$P_{\rm PMPS}$\tablenotemark{b}} &
\colhead{DM/$P_{\rm PALFA}$\tablenotemark{b}}
}
\startdata 
Observed &  20   & $>15$ &   8 & 33 \\
Model A  &  20   &  40   &  14 & 56 \\
Model B  &  20   &  15   &   9 &  7 \\
Model C  &  20   &  37   &  11 &113 \\
Model D  &  20   &  42   &  16 & 27 
\enddata

\tablecomments{The four models and their parameters are 
described in the text in \S3.}

\tablenotetext{a}{Number of detectable MSPs.}

\tablenotetext{b}{Median DM/$P$ value of detectable MSPs.}

\end{deluxetable}


\end{document}